\documentclass{emulateapj} 
\usepackage{apjfonts} 

\def\gsim{\mathrel{\rlap{\lower2pt\hbox{\hskip0pt\small$\sim$}}
\raise2pt\hbox{\small $>$}}}
\def\lsim{\mathrel{\rlap{\lower2pt\hbox{\hskip0pt\small$\sim$}}
\raise2pt\hbox{\small $<$}}}     \def\bq{\begin{equation}}
\def\eq{\end{equation}}

\newcommand{\beq}{\begin{equation}}
\newcommand{\eeq}{\end{equation}}

\shorttitle{Reionization by the First Galaxies} \shortauthors{ALVAREZ, FINLATOR \& TRENTI}

\begin{document}

\title{Constraints on the Ionizing Efficiency of the First Galaxies}

\author{Marcelo~A.~Alvarez\altaffilmark{1}, Kristian Finlator\altaffilmark{2} and
Michele Trenti\altaffilmark{3}}

\altaffiltext{1}{Canadian Institute for Theoretical Astrophysics,
University of Toronto, 60 St.George St., Toronto, ON M5S 3H8, Canada}
\altaffiltext{2}{Department of Physics, University of California Santa
  Barbara, Santa Barbara, CA 93106, USA}
\altaffiltext{3}{Institute of Astronomy, University of Cambridge,
  Madingley Road, Cambridge, CB3 0HA, United Kingdom}
\email{malvarez@cita.utoronto.ca}

\begin{abstract}

Observations of the Lyman-$\alpha$ forest and of high-redshift
galaxies at $z\sim 6-10$ imply that there were just
enough photons to maintain the universe in an ionized state at $z\sim
5-6$, indicating a ``photon-starved'' end to reionization.  The ionizing
emissivity must have been larger at earlier times in order to yield the
extended reionization history implied by the electron scattering optical 
depth constraint from WMAP.
Here we address the possibility that a faint population of
galaxies with host halo masses of $\sim 10^{8-9}M_\odot$ dominated
the ionizing photon budget at redshifts $z\gsim 9$, due to their much higher
escape fractions. Such faint, early galaxies, would not have formed in
ionized regions due to suppression by heating from the UV background,
and would therefore not contribute to the ionizing background at
$z\lsim 6$, after reionization is complete. Our model matches: (1) the
low escape fractions observed for high-redshift galaxies, (2) the WMAP
constraint of $\tau_{\rm es}\sim 0.09$, (3) the low values for the UVB
at $z<6$, and (4) the observed star formation rate density inferred
from Lyman-break galaxies.  A top heavy IMF from Pop III stars is
not required in this scenario. We compare our model to recent ones in the
literature that were forced to introduce an escape fraction that
increases strongly towards high redshift, and show that a similar
evolution occurs naturally if low mass galaxies possess high escape 
fractions.  

\end{abstract}


\section{Introduction}

A fundamental question in the theory of reionization is:
{\em what were the ionizing sources?} While it was
realized several decades ago that the observed quasar population was
insufficient \citep{shapiro/giroux:1987}, the actual sources have
eluded detection. The main
constraints on reionization models are the Gunn-Peterson trough
\citep{gunn/peterson:1965} in the spectra of high redshift quasars at a redshift of $z\sim 6$
\citep[e.g.,][]{fan/etal:2002,willott/etal:2007}, and the 
electron scattering optical depth, $\tau_{\rm es}$, with the
latest results indicating that the universe was substantially ionized by
$z\sim 10$ \citep{komatsu/etal:2011}.  Models adhering to these
constraints predict extended reionization histories
\citep[e.g.,][]{haiman/holder:2003, cen:2003, wyithe/loeb:2003,
  ricotti/ostriker:2004, choudhury/ferrara:2006, fan/etal:2006,
  haiman/bryan:2006, iliev/etal:2007,
  kuhlen/faucher-giguere:2012, haardt/madau:2012, ahn/etal:2012}.

The ionizing emissivity observed at $z\sim$4--6 
is too low to account for the observed $\tau_{\rm es}$.
~\citet{bolton/haehnelt:2007a} found that extrpolating the ionizing emissivity
inferred from the Lyman-$\alpha$ forest at $z\sim$4--6 to higher
redshifts would underproduce $\tau_{\rm es}$. Likewise, observations of 
``Lyman-break galaxies'' (LBGs) at $z>6$ indicate a rapidly declining 
star formation rate in the brightest galaxies toward high redshift
\citep{oesch/etal:2012}, consistent with evolution of the
dark-matter halo mass function \citep{trenti/etal:2010}.  The ionizing
emissivity from these galaxies can reionize the universe by $z\sim 6$, under
fairly generous assumptions regarding uncertainties such as the ionizing
escape fraction, the intergalactic medium temperature, and the abundance 
of galaxies fainter than the current detection
limit~\citep{trenti/etal:2010,bouwens/etal:2011}.  Thus, both
observations  of the high-redshift galaxy population {\em and} of the
Lyman-$\alpha$  forest indicate a photon-starved end to reionization.  

Reconciling the ionizing emissivity observed at $z\sim 4-6$ with the
$\tau_{\rm es}$ value has been the subject of recent work.
\citet{kuhlen/faucher-giguere:2012} presented a model that matches
both a photon-starved end to reionization and the WMAP result. In
their best-fitting model, the escape fraction, $f_{\rm esc}$, varies
with redshift, while the relation between halo mass and ionizing
luminosity is extrapolated from abundance matching at lower
redshift. The ``minimal reionization model'' adopted by \citet[][--
HM12 hereafter]{haardt/madau:2012} also assumes $f_{\rm esc}$
increases with redshift. \citet{fontanot/etal:2012} find that either
an increasing $f_{\rm esc}$ towards high redshift and/or towards
fainter galaxies is required. \citet{shull/etal:2012} argue instead
for an end of reionization of $z\approx7$ and a $\tau_{\rm es}$ at the
low-end of the WMAP allowed range.  

Here we investigate the scenario in which low mass galaxies have a
much higher $f_{\rm esc}$ than more massive ones, but do not survive past
reionizion. We separate halos into two categories based on their mass.
Low-mass halos are subject to complete suppression by a photoionizing
background and are called  ``photosuppressible'', while more massive
halos are not sensitive to an ionizing background. Under these
circumstances, photosuppression
effects naturally extend the reionization process \citep[e.g.,][]{haiman/holder:2003,wyithe/loeb:2003,choudhury/ferrara:2006,
iliev/etal:2007}.  We further assume
that low-mass halos have similar star formation efficiencies to their
high-mass counterparts, but have much higher escape fractions, of
order unity.  Before proceeding, we first discuss the two main
physical motivations for our model.   

First, theoretical models and observations indicate the ionizing
photon $f_{\rm esc}$ should vary with halo mass. Star formation in the
lowest-mass galaxies was likely bursty, taking place in a highly
irregular gas distributions more resembling massive star forming
regions than rotationally supported disks. As illustrated by the
numerical simulations presented by \citet{wise/cen:2009}, stars should
be able to carve out channels through which radiation can escape. For
higher mass halos, the escape fraction should only reach a few per
cent, because most of the ionizing  radiation is produced deep within
the galaxy's disk and absorbed
\citep{gnedin/etal:2008,razoumov/sommer-larsen:2010}.  In support of
this picture, observations  at $z\sim3$ indicate that Lyman Break
galaxies possess  low ionizing escape fractions
\citep[$<10\%$;][]{shapley/etal:2006,siana/etal:2010} while fainter
galaxies show escape fractions that grow comparable to
unity~\citep{iwata/etal:2009,nestor/etal:2011}.

Second, photoionization heating has a negative feedback effect on star
formation in dwarf galaxies
\citep{efstathiou:1992}. \citet{shapiro/etal:1994} investigated the
linear growth of gas perturbations, finding a filtering scale that
lags behind the instantaneous Jeans scale, and \citet{gnedin/hui:1998}
presented convenient analytical formulae for the filtering scale.
Subsequent coupled radiation-hydrodynamical simulations confirmed the
negative feedback effect of photoionization heating
\citep{thoul/weinberg:1996, gnedin:2000b,
dijkstra/etal:2004,okamoto/etal:2008}. The latest simulations by
\citet{finlator/etal:2011} indicate a suppression scale of a few times
$10^9~M_\odot$. Photoionization heating will have a significant
negative effect on star formation in the faintest, most abundant,
early galaxies.   

In this paper we attempt to find the simplest possible reionization
model that satisfies current constraints. An important simplifying
assumption we have made is the same halo mass threshold for the escape
fraction {\em and} photosuppression -- in our fiducial model, halos
with masses below $2\times 10^9 M_\odot$ have both high escape
fractions and are subject to photosuppression. A common mass scale
is physically plausible, however, given that it is the same mechanism
(photoionization feedback) which is likely to be responsible both for
higher escape fractions and suppression in low-mass halos.

We present our model in \S 2, results in \S 3, and
a discussion in \S 4. All distances and densities are in comoving
units, and we adopt cosmological parameters consistent with the latest
WMAP constraints \citep{komatsu/etal:2011},
$(\Omega_m,\Omega_\Lambda,\Omega_b,h,\sigma_8,n_s)=(0.275,0.725,0.046,0.7,0.82,0.97)$.

\begin{figure*}
\begin{centering}
\hspace{1.cm}\includegraphics[width=0.9\textwidth]{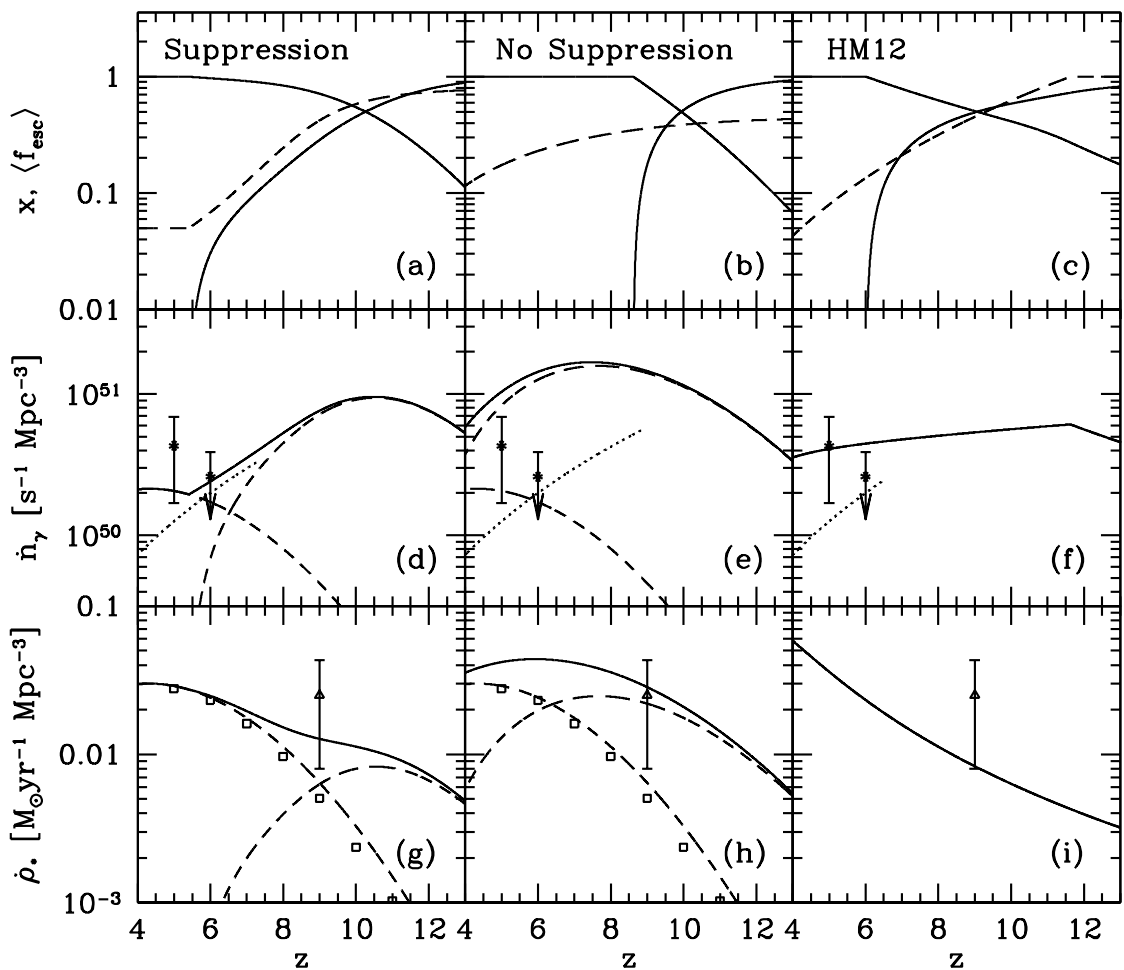}
\vspace{0.3cm}
\end{centering}
\caption{
{\em top}: Ionized fraction $x(z)$ and neutral fraction $1-x(z)$.  The
dashed line shows the mean escape
fraction {\em middle}: Comoving ionizing
photon emissivity $\dot{n}_\gamma(z)$ (solid), as well as the
contribution from halos in each of our two mass ranges (dashed).
Solid points with errorbars show
measurements from the Lyman-alpha forest at $z=$5--6, as compiled
by \citet[][see their Table 2 and references
therein]{kuhlen/faucher-giguere:2012}. Dotted lines indicate the
emissivity required to keep the universe ionized,
$\dot{n}_\gamma=n_0/\langle t_{\rm rec}\rangle$, and are only plotted
for redshifts such that $x>0.9$.
{\em bottom}: Comoving SFRD, $\dot{\rho}_*(z)$,
for each of the models (solid), as well as the contribution from halos
in each of our two mass ranges (dashed). The triangular point with
errorbars is the value obtained from GRB observations at $z\sim 9$ by
\citet{robertson/ellis:2012}.  Square points indicate a fit to the 
observed SFRD for halos with $M>2\times 10^9 M_\odot$, and its
extrapolation to $z\geq8$ \citep{trenti/etal:2010}. 
}
\vspace{0.5cm}
\label{fig1}
\end{figure*}

\section{Model}

We obtain the hydrogen ionized fraction, $x(z)$, using the following
relationship \citep[e.g.,][]{shapiro/giroux:1987, haardt/madau:2012}:  
\beq
\dot{x} = \frac{\dot{n}_\gamma}{n_0}-\frac{x}{\langle t_{\rm rec}\rangle},
\label{history}
\eeq
where $\dot{n}_\gamma$ is the comoving ionizing photon emissivity,
$n_0=n_{\rm H}+n_{\rm He}$ is the comoving number density of hydrogen
and helium nuclei, and the recombination time is $\langle t_{\rm
 rec}\rangle^{-1}=n_0\alpha_Bc_l(z)(1+z)^3$. We use the clumping factor
$c_l=C_{100}$ obtained by \citet{pawlik/etal:2009} for their $z_{\rm
 reion}\sim 10.5$ case, $c_l=e^{-0.28z+3.59}+1$ for $z>10$ and
$c_l=3.2$ for $z<10$, consistent with the
transmissivity and mean free path reported in the literature 
\citep{songaila/cowie:2010,kuhlen/faucher-giguere:2012}.
Helium is singly ionized along with hydrogen, with He II reionization
occurring later, at $z=3$.  

The ionizing emissivity is related to the star formation rate density (SFRD)
via 
\beq
\dot{n}_\gamma=f_{\rm esc,1}f_\gamma\dot{\rho}_{*,1}+f_{\rm esc,2}f_\gamma\dot{\rho}_{*,2}
\label{ndot}
\eeq
where $f_{\rm esc,i}$ corresponds to type $i$ halos, $f_\gamma$ is the number of
ionizing photons produced per stellar mass formed, and $\dot{\rho}_{*,i}$ is
the SFRD of type $i$ halos. Type 1 (photosuppressible)
halos can only host sources of ionizing radiation if they are in
neutral regions, while more massive type 2 halos can host sources even
after reionization. 

We use
a Salpeter IMF, with $M_{\rm min}=0.1 M_\odot$, $M_{\rm max}=100
M_\odot$. At solar metallicity with this IMF, $f_\gamma\simeq 3\times
10^{60} M_\odot^{-1}$ \citep[e.g.,][and references
therein]{haardt/madau:2012}, while for a metal poor population with
$Z=0.02Z_\odot$, $f_\gamma\simeq 5\times 10^{60} M_\odot^{-1}$, and
for the $Z=0$ case $f_\gamma\simeq 7\times 10^{60} M_\odot^{-1}$
\citep{schaerer:2002}. 
We assume that the ionizing photons are emitted
instantaneously, which is an excellent approximation given the
lifetime of the massive stars that produce the bulk of the ionizing radiation.

The SFRD is given by 
\beq
\dot{\rho}_*=\dot{\rho}_{*,1}+\dot{\rho}_{*,2}=\rho_0\frac{\Omega_b}{\Omega_m}\left[
(1-x)\epsilon_{*,1}\dot{f}_1+\epsilon_{*,2}\dot{f}_2
\right],
\label{sfr}
\eeq
where $\epsilon_{*,i}$ is the fraction of baryons converted to stars
in halos in the mass range $i$ with collapsed fraction $f_i$. 
The $(1-x)$ factor in the first term accounts
for the suppression of star formation in ionized regions for
photosuppressible 
halos.  We will use $M_1$ and $M_2$ to indicate the lower mass limits
of our two halo mass ranges.  Equations (\ref{history}), (\ref{ndot}),
and (\ref{sfr}) are solved simultaneously to obtain the ionized
fraction, ionizing emissivity, and star formation rate density as a
function of redshift.

The electron scattering optical
depth from reionization is  
\beq
\tau_{\rm es}=\frac{3H_0\Omega_bc\sigma_T}{8\pi Gm_p}\int_0^\infty
\frac{x(z)(1+z)^2(1-Y+N_{He}(z)Y/4)}{\sqrt{\Omega_m(1+z)^3+1-\Omega_m}}dz, 
\eeq
where $Y\simeq 0.24$ is the helium abundance and $N_{\rm He}(z)$ is
the number of times helium is ionized -- we use $N_{\rm He}=2$ for
$z<3$ and $N_{\rm He}=1$ for $z>3$.   

\section{Results}

Shown in Figure \ref{fig1} are our results for our three models.
The first two, ``suppression'' and ``no suppression'', were
obtained using the model in \S 2. In both cases,
$M_1=10^{8}M_\odot$ (roughly the mass at which atomic cooling is
efficient), $M_2=2\times 10^9 M_\odot$ (the mass at above which star
formation can proceed in ionized regions), $f_{\rm esc,2}=0.05$, 
and $\epsilon_{*,1}=\epsilon_{*,2}=0.03$. In the no suppression model,
we removed the $1-x$ factor on the right hand side
of equation (\ref{sfr}). We restricted $\epsilon_{*,1}\leq \epsilon_{*,2}$, since internal feedback effects are
stronger at lower masses, and $\epsilon_{*,2}=0.03$ was chosen so as
to match the overall efficiency found of halos with $M>2\times 10^9
M_\odot$ in the \citet{trenti/etal:2010} ICLF model (see Figure 1 and
discussion below). We chose $f_{\rm esc,2}=0.05$ to be consistent with
the relatively low escape fractions inferred for observed LBGs.
We varied the remaining parameter, $f_{\rm esc,1}$, in both cases to
obtain $\tau_{\rm es}= 0.086$, so that $f_{\rm esc,1}=0.8$ and $f_{\rm
  esc,1}=0.45$ in the cases with and without suppression,
respectively.  

The last model shown is our implementation of the ``minimal
reionization'' model of HM12, with a constant value of $f_\gamma$.
In their implementation, the metallicity is assumed to follow $Z(z)=Z_\odot
10^{-0.15z}$, which corresponds to a value of $f_\gamma\sim 5\times
10^{60} M_\odot^{-1}$ at $z\sim 11$, decreasing at lower
redshifts. Since such an evolution is degenerate with a slight
change  $f_{\rm esc}(z)$, we do not expect this to be important. 
We used their fit for $\dot{\rho}_*(z)$,
which was extrapolated beyond $z\sim 7$ from the observed data
(overproducing $\dot{\rho}_*(z=10)$ according to
\citealt{oesch/etal:2012}), and an evolving escape fraction, $f_{\rm
  esc}(z)=1.8\times 10^{-4}(1+z)^{3.4}$. For the HM12 model, we also
obtained $\tau_{\rm es}=0.086$.  In all three cases we used
$f_\gamma=4.5\times 10^{60} M_\odot^{-1}$. 

\subsection{With suppression}

As seen in  panel (a) of Figure 1, the ionized fraction $x$ passes
50\% at $z\sim10$ and 90\% at $\sim7$.  Also shown in panel (a) is the
mean emission-weighted escape fraction, defined by  
\beq
\langle f_{\rm
  esc}\rangle(z)=\frac{\dot{n}_\gamma(z)}{f_\gamma\dot{\rho}_*(z)}.
\label{fesc} 
\eeq 
The escape fraction rises from $\sim 0.05$ at $z<6$, to $\sim 0.8$ at
$z=11$. This evolution is entirely due to the evolution in the
underlying source population -- as the universe becomes more neutral,
more ionizing photons come from photosuppressible halos, which have
much higher escape fractions.   

The ionizing emissivity is shown in panel (d), as well as constraints from 
table 2 of \citet{kuhlen/faucher-giguere:2012}, derived by combining
measurements of the mean free path
\citep[e.g.,][]{songaila/cowie:2010} with the transmissivity of the
Lyman-$\alpha$ forest \citep[e.g.,][]{bolton/haehnelt:2007a}.
It reaches a peak of
$\dot{n}_\gamma\sim 10^{51}{\rm s}^{-1}{\rm Mpc}^{-3}$ at $z\sim
11$. The upturn at $z\sim 5$ is due to the rise of higher mass
sources, in particular from halos above $2\times 10^{9}M_\odot$.
The main sources of reionization in this model do not contribute
substantially to the emissivity at $z<6$, consistent with {\em both}
an early reionization that satisfies the WMAP optical depth, {\em and}
a photon-starved end to reionization. The star formation rate density
rises from 
$\dot{\rho}_*\sim 10^{-3}M_\odot{\rm  yr}^{-1}{\rm Mpc}^{-3}$ at
$z=16$, plateaus at $\sim 0.01$ over $z\sim 8-12$, increasing at
$z<12$. (panel g).  Shown also are values from the ICLF model of
\citet{trenti/etal:2010} integrated down to a minimum halo mass of
$2\times 10^9 M_\odot$ -- our model matches the predicted evolution of
the SFRD of Lyman-break galaxies.  By $z\sim 13$, star formation
occurring in suppressible halos accounts for over 90 per cent of the
total SFRD.  

\subsection{No suppression}

As seen in panel (b), overlap occurs earlier in the model
without suppression, at $z\sim
8.5$ (as compared to $z\sim 5.5$ in the suppression model). 
The post-reionization ionizing emissivity is higher without
suppression, $\dot{n}_\gamma>10^{51}{\rm s}^{-1}{\rm Mpc}^{-3}$ for
$z\gsim 5$ (panel e). This violates inferences from the Lyman-$\alpha$
forest at $z\sim 5-6$, which indicate values at those redshifts of
$\sim 2-4\times 10^{50}{\rm s}^{-1}{\rm Mpc}^{-3}$ \citep[data points
in Figure
\ref{fig1};][]{bolton/haehnelt:2007a,kuhlen/faucher-giguere:2012}, and
would be difficult to reconcile with the appearance of the
Gunn-Peterson trough at $z\gsim 6$
\citep[e.g.,][]{fan/etal:2002,willott/etal:2007}. We were unable to
find a model with $\epsilon_{*,1}\leq \epsilon_{*,2}$ and $f_{\rm
  esc,2}\leq 0.05$ that simultaneously satisfies the constraints and
does not include suppression.

\subsection{Minimal reionization model of HM12}

In Figure 1, we show results from our implementation of the
``minimal reionization  model'' of HM12. Shown in panel (c) is the
escape fraction, $f_{\rm esc}\propto (1+z)^{3.4}$. 
Since $\dot{n}_\gamma\propto f_{\rm esc}\dot{\rho}_*$, the drop in the 
SFRD towards higher redshift in panel (i) is roughly canceled
by the increasing $f_{\rm esc}$, so that $\dot{n}_\gamma$ is nearly
flat (panel f). Notice that scenarios where the
ionizing emissivity is flat struggle to reproduce the upper
limit to the emissivity at $z=6$.  By contrast, the suppression model 
naturally accommodates this limit with an increasing
contribution from fainter sources at higher redshifts (panel d). However,
the upper limit at $z=6$
($\dot{n}_\gamma<2.6\times 10^{50}{\rm s}^{-1}{\rm Mpc}^{-3}$) could
be higher, as it is based on combining inferred photoionization rates
\citep{bolton/haehnelt:2007a} with highly-uncertain estimates of the
mean free path at $z=6$ \citep{songaila/cowie:2010}. A
lower mean free path than assumed by
\citet{kuhlen/faucher-giguere:2012}, which was based upon a fit which
lies above the $z=5.7$ data point from \citet{songaila/cowie:2010},
would result in a higher upper limit. 

Both the HM12 model and our suppression model exhibit a similarly
strong evolution in the mean escape fraction, with $f_{\rm esc}$ of
order unity for $z>10$, and $f_{\rm esc}\sim 0.05$ at $z\sim 5$. In
the suppression model, this is a natural outcome of the assumption
that the low-mass halos have high escape fractions and dominate the
ionizing photon budget early ($f_{\rm esc,1}=0.8$), but by the end of
reionization only the more massive halos host stars, and the escape
fraction is about a factor of 15 lower ($f_{\rm
esc,2}=0.05$). Suppression can therefore be viewed as a physical
explanation for the strongly evolving escape fraction needed to
satisfy the observational constraints.

Comparing the SFRDs assumed by the HM12 and the suppression model
reveals that they agree to within a factor of 2 for all $z=$5--13.
In detail, the SFRD in the HM12 model uses an
extrapolation to $z>8$ of a fitting function constrained at $z<8$:
\beq
\dot{\rho}_* = \left[6.9\times
  10^{-3}+0.14\frac{(z/2.2)^{1.5}}{1+(z/2.7)^{4.1}}\right]M_\odot{\rm
    yr}^{-1}{\rm Mpc}^{-3}.
\eeq
The suppression model predicts 
that the mass range dominating star formation transitions smoothly
from photosuppressible halos to more massive ones around $z=$8--10.
In fact, comparing the dashed curves in panel (g) with 
the SFRD in panel (i), the HM12 model depends on either the increasing role
of sources that cannot be detected in the post-reionization Universe
(including photosuppressible halos), or a  strongly increasing star
formation efficiency in LBGs.  

\section{Discussion and Observational Implications}\label{sec:discussion}

In this letter we presented a reionization model that can
simultaneously satisfy all the current observational constraints on
hydrogen reionization (Lyman-$\alpha$ forest at $z\sim5-6$, SFRD from
Lyman-break galaxy surveys at $z\sim6-10$, low escape fractions at
$z\lsim 6$, and WMAP $\tau_{es}$). Our model predicts that the
majority of the ionizing photons at $z\gtrsim9$ are produced in faint
galaxies hosted in photosuppressible dark matter halos with $M\sim
10^{8.5}~M_{\odot}$, undetectable in current surveys
\citep{trenti/etal:2010,munoz/loeb:2011,kuhlen/faucher-giguere:2012}. Once
reionization is completed, the star formation in these halos is
suppressed by radiative feedback, so that these sources do not
contribute to the ionizing emissivity at lower redshift. Therefore, we
predict that the sources of reionization are primarily galaxies with
$L<0.06L_*(z=3)$, too faint to be detected by even the deepest HST
observations.

A key prediction is that the SFRD is higher than
that inferred from Lyman-break galaxy surveys, by a factor of
$\sim 2$ at $z=9$, increasing to a factor of $\sim 10$ at
$z=12$. Indeed, alternate tracers of star formation indicate that this
is likely. Estimates of $\dot{\rho}_*$ from the
GRB rate measure $\dot{\rho}_*\sim
0.025_{-0.017}^{+0.018}~\mathrm{M_{\sun} yr^{-1} Mpc^{-3}}$ at $
z\sim9$ (\citealp{robertson/ellis:2012}; see also
\citealp{kistler/etal:2009}), one order of magnitude higher than the
LBG surveys estimate. In addition, the non-detection of GRB host
galaxies at $z>5$ in ultradeep HST images provides an independent
confirmation that the majority of star formation at those redshifts is
happening in galaxies fainter than current observational limits
\citep{trenti/etal:2012}. Finally, the latest determination of the
galaxy luminosity function at $z\sim8$ has a very steep faint-end
slope ($\alpha=-1.98\pm0.2$) at $M_{\rm AB}\sim-17.7$ , implying a
logarithmically divergent contribution to the SFRD from fainter,
unseen galaxies \citep{bradley/etal:2012}. All these observations are
naturally explained in the framework presented here based on
photosuppressible halos providing most of the reionizing photons.  

It is useful to compare the SFRD and escape fractions we obtain to
that of the HM12 model, which is based on the SFRD in galaxies with
$L>0.06L_*$ at $z=2-7$. In our framework the sources of reionization
at $z\sim10$ will remain unseen in near-future galaxy surveys at
$z\sim10$ -- our photosuppressible halos correspond to $-15\lsim
M_{\rm AB}\lsim -10$ over $z\simeq 6-12$. If the SFRD in the HM12 model
comes from galaxies of the same UV luminosities as
at lower redshifts, their extrapolation to $z=10$ is likely
too high and marginally inconsistent with the strong decline in the
LBG population between $z=8$ and $z=10$ observed by \citet[][see their
Fig. 8]{oesch/etal:2012}.

Even assuming a SFRD due to  bright galaxies that is not declining as
quickly as suggested by the LBG observations, it is still necessary to
assume an escape fraction that is a strongly increasing function of
redshift in order to satisfy the WMAP $\tau_{\rm es}$ constraint. Such
an evolution is naturally explained in our model, where the average
escape fraction increases to high redshift simply because there are
more low-mass halos forming in the unheated regions.

The high escape fractions for faint galaxies favored in our model
should have a unique signature in the cosmic infrared background
\citep{fernandez/etal:2012}. This is because the Lyman-$\alpha$
emission per escaping ionizing photon is inversely proportional to the
escape fraction. A larger escape fraction for a fixed ionization
history will result in a lower overall flux and fluctuation amplitude. 

There are several important caveats. First, we have assumed
that the star formation efficiency in photosuppressible halos
is the same as more massive ones. Simulations by \cite{wise/etal:2012b} that
include the effects of radiation pressure, supernova feedback, and
metal-line cooling, find $\dot{M}_*\sim 2\times 10^{-2}M_\odot{\rm
yr}^{-1}$ for a $2\times 10^8M_\odot$ halo at $z=8$, implying that
$\sim 0.03$ of the gas accreted after reaching the
atomic cooling limit is converted to stars, consistent with our choice
of $\epsilon_{*,1}=0.03$. The simulations of
\citet{finlator/etal:2011}, on the other hand, exhibit a much lower
star formation rate at a similar mass of a few times $10^8
M_\odot$. Although theoretical predictions have not converged,
$\epsilon_{*,1}\simeq \epsilon_{*,2}$ seems plausible.

In addition, we have assumed that the suppression operates
instantaneously within H~II regions. Although this is an
oversimplification, it illustrates the basic trends resulting from
high escape fractions and suppression in low-mass halos. Detailed
numerical simulations that self-consistently include both external and
internal radiative feedback during reionization will be required to
address these issues more accurately. 

Finally, we note that the best-fitting value for the electron
scattering optical depth from WMAP, $\tau_{\rm es}=0.088$, has
1-$\sigma$ error bars of of $\sim 0.015$, indicating that the true
value could be as low as $\tau_{\rm es}\sim 0.06$, at 2-$\sigma$. In
our model with suppression, $\tau\sim 0.06$ can be achieved with a
lower escape fraction for low mass halos, $f_{\rm esc,1}\sim
0.2$. This is still however a factor of four times higher than that
for the larger mass halos, $f_{\rm esc}=0.05$. 
Upcoming measurements with the Planck satellite will be
essential in tightening constraints on $\tau_{\rm es}$ and therefore
on the nature of the first galaxies.  

\vspace{-0.2cm}
\acknowledgments{We thank T. Abel and R. Wechsler for helpful
  discussions, M.~McQuinn and S.~P.~Oh for comments on an earlier
  draft, the anonymous referee, and the Kavli Institute for Theoretical
  Physics, Santa Barbara, for their hospitality during the workshop
  ``First Galaxies and Faint Dwarfs: Clues to the Small Scale
  Structure of Cold Dark Matter", where this work was initiated. This
  research was supported in part by the National Science Foundation
  under Grant No. NSF PHY11-25915.}       


\end{document}